\documentclass[aps,prb,twocolumn,superscriptaddress,groupedaddress,showpacs]{revtex4}
\usepackage{graphicx,xcolor}



\begin{document}

\title{Mean-field embedding of the dual fermion approach for correlated electron systems}

\author{S.-X. Yang}
\email{yangphysics@gmail.com}
\affiliation{Department of Physics and Astronomy, Louisiana State University, Baton Rouge, Louisiana 70803, USA}
\affiliation{Center for Computation and Technology, Louisiana State University, Baton Rouge, Louisiana 70803, USA}
\author{H. Terletska}
\affiliation{Condensed Matter Physics and Materials Science Department,
Brookhaven National Laboratory, Upton, New York 11973, USA }
\author{Z. Y. Meng}
\affiliation{Department of Physics and Astronomy, Louisiana State University, Baton Rouge, Louisiana 70803, USA}
\affiliation{Center for Computation and Technology, Louisiana State University, Baton Rouge, Louisiana 70803, USA}
\author{J.\ Moreno}
\affiliation{Department of Physics and Astronomy, Louisiana State University, Baton Rouge, Louisiana 70803, USA}
\affiliation{Center for Computation and Technology, Louisiana State University, Baton Rouge, Louisiana 70803, USA}
\author{M.\ Jarrell}
\affiliation{Department of Physics and Astronomy, Louisiana State University, Baton Rouge, Louisiana 70803, USA}
\affiliation{Center for Computation and Technology, Louisiana State University, Baton Rouge, Louisiana 70803, USA}
\date{\today}

\begin{abstract}
To reduce the rapidly growing computational cost of the dual fermion lattice calculation with increasing system size,
we introduce two embedding schemes. One is the real fermion embedding, and the other is the dual fermion embedding. 
Our numerical tests show that the real fermion and dual fermion embedding approaches converge to essentially the same result. 
The application on the Anderson disorder and Hubbard models shows that these embedding algorithms converge more quickly 
with system size as compared to the conventional dual fermion method, for the calculation of both single-particle and 
two-particle quantities.
\end{abstract}

\pacs{02.70.-c, 71.27.+a, 71.10.Fd, 71.30.+h}

\maketitle

\section{Introduction}

Mean-field methods like the Coherent Potential Approximation (CPA)\cite{p_soven_67,Elliott-cpa} and the Dynamical Mean-Field 
Theory (DMFT)\cite{w_metzner_89a,e_mullerhartmann_89a,t_pruschke_95,Georges_DMFT} are widely applied to the study of 
disordered and correlated 
materials.  By construction, these methods are single-site mean-field approximations, where the real lattice is replaced by an 
impurity placed in a local (momentum-independent) effective medium. As single-site approximations, both the CPA and DMFT fail 
to take into account nonlocal inter-site correlations and fluctuations of the medium, which are found to be important 
in many materials with nonlocal order parameters or strong inter-site correlations.

To systematically incorporate such nonlocal corrections to these mean-field approaches, cluster extensions of the DMFT and CPA,
such as the Dynamical Cluster Approximation (DCA)\cite{m_hettler_98a,m_hettler_00a,m_jarrell_01a,Maier05} have been 
developed.  Here a finite size periodic cluster of several lattice sites is placed in a self-consistently determined 
effective medium, which now acquires cluster-resolved momentum-dependence. The embedding is achieved by coarse graining the
lattice problem in momentum space.  Such a cluster embedding allows for explicit treatment of short-range correlations and 
non-local order parameters within the cluster size, while the longer length scale physics is still described at the 
mean-field level.  The cluster may be  solved with numerically exact methods such as quantum Monte Carlo or exact 
diagonalization.  Unfortunately, these quantum cluster methods are limited by the computation effort needed for the 
cluster solvers.  Exact diagonalization has an exponential scaling in cluster size and quantum Monte Carlo is plagued 
by the fermion sign problem ~\cite{sign_problem1}.

To address such an exponential scaling, methods have been developed which map the lattice problem onto an impurity 
self-consistently embedded in a correlated lattice problem\cite{Janis_parquet,Toschi_07,c_slezak_06b,Rubtsov08}.  Here, 
local correlations are treated on the impurity, while nonlocal correlations are incorporated on the lattice via a 
diagrammatic perturbation expansion around the DMFT solution.  If a QMC method is used to solve the impurity problem, 
and if the impurity is small enough that the fermion sign problem is absent or controllable, then these methods scale 
algebraically in the lattice size.  The dual fermion\cite{Rubtsov08} approach is perhaps the most elegant of these 
methods since here the mapping to an embedded impurity is apparently exact, provided that the lattice perturbation 
theory can be solved to all orders.  

One of the practical constraints in the implementation of the dual fermion method is that its computational complexity 
increases with the lattice size.  The lattice size should be large enough to represent a thermodynamic limit, but 
this can make the diagrammatic calculation on the lattice computationally expensive.  Becasue of such limitation
the dual fermion approach has been applied mostly to one- and two-dimensional systems, and not yet to three-dimensional 
systems. To overcome this issue, we introduce an extension of the dual fermion method to include a third length scale 
introduced to reduce 
the complexity involved in the treatment of the correlations at the intermediate length scale.  Here, using ideas 
from the DCA, the dual fermion lattice is replaced by a DCA cluster embedded in a self-consistently determined 
effective medium.  Two algorithms are presented, one employs the DCA coarse graining on the real fermion lattice 
and the other on the dual fermion lattice.  We find that the latter approach is more efficient and that this 
modification dramatically improves the convergence of the dual fermion method with system size and enables the use  
of higher order approximations for the diagrammatic solution to the cluster problem.

This paper is organized as follows. In Section II after reviewing the dual fermion algorithm, we provide a detailed description 
of the two proposed embedding schemes. Then in section III, to test our methods we first apply them to the one-dimensional 
Anderson disorder model. And in section IV, we demonstrate its application on the two-dimensional Hubbard model.  
The numerical results show a superior convergence of our embedding schemes as 
compared to the conventional dual fermion algorithm as a function of the lattice size. Section V summarizes and concludes 
the paper.

\section{Formalism}

\subsection{Dual fermion mapping}

To derive the dual fermion formalism for either interacting~\cite{Rubtsov08} and disordered systems~\cite{h_terletska_13, df_afk}, 
we start from the lattice action
\begin{equation}
S[c,c^*] = -\sum_{\omega,{\bf k},\sigma}(i\omega + \mu - \epsilon_{\bf k}) 
c^*_{\omega,{\bf k},\sigma} c_{\omega,{\bf k},\sigma}
+ \sum_i S_{loc}[c_i,c^*_i],
\label{eq:action}
\end{equation}
where $S_{loc}[c_{i}^{*},c_{i}]$ is the local part of the action (e.g., a Hubbard
interaction term or a local disorder potential), $c_{i}^{*}$ and $c_{i}$ are Grassmann numbers 
corresponding to creation and annihilation operators on the lattice, $\mu$ is the chemical potential, 
$\epsilon_{\bf k}$ is the lattice bare dispersion, and $\omega=(2n+1)\pi T$ are the Matsubara frequencies.
For interacting systems, this action is used to calculate the partition function~\cite{Rubtsov08} while for 
disordered systems the replica method may be used to directly calculate the Green functions~\cite{h_terletska_13, df_afk}.  
Then to express this action in terms of single impurity problem 
\begin{equation}
S_{imp}[c_i,c^*_i]= -\sum_{\omega,\sigma}\mathcal{G}(iw)^{-1} 
c^*_{\omega,i,\sigma} c_{\omega,i,\sigma}
+ S_{loc}[c_i,c^*_i]
\label{eq:imp_action}
\end{equation}
we rewrite Eq.~\ref{eq:action} as
\begin{equation}
S[c,c^*] = -\sum_{\omega,{\bf k},\sigma}(\Delta_w - \epsilon_{\bf k}) 
c^*_{\omega,{\bf k},\sigma} c_{\omega,{\bf k},\sigma}
+ \sum_i S_{imp}[c_i,c^*_i],
\label{eq:action2}
\end{equation}
here the impurity-excluded (bath) Green function is defined as $\mathcal{G}(iw) \equiv (iw + \mu - \Delta_w)^{-1}$ 
and $\Delta_w$ is the hybridization function between the impurity and the effective medium.
By introducing the auxiliary (dual fermion) degrees of freedom $f^*_{\omega {\bf k} \sigma}, f_{\omega {\bf k} \sigma}$ 
via a Hubbard-Stratonovich transformation of the first term in Eq.~\ref{eq:action2}, and then integrating 
out the real fermion degrees of freedom~\cite{Rubtsov08, hartmut_thesis} 
(see Appendix A in Ref.~\onlinecite{hartmut_thesis} for a detailed derivation),
 we end up with the following dual fermion 
action
\begin{eqnarray}
S_d[f^{*},f] = -\sum_{{\bf k} \omega \sigma}f^*_{\omega {\bf k} \sigma} G^0_d({\bf k},i\omega)^{-1}f_{\omega {\bf k} \sigma} 
 +\sum_{i} V[f^*_i,f_i ],
\label{eq:DFaction}
\end{eqnarray}
where $G^0_d$ is the bare dual Green function defined as the difference between the DMFT
and/or CPA lattice Green function 
$G_{lat}$ and the impurity Green function $G_{imp}$, i.e.,
\begin{equation}
G^0_{d}({\bf k},i\omega) = G_{lat}({\bf k},i\omega) - G_{imp}(i\omega).
\label{Eq:dual-Green}
\end{equation}
The dual fermion potential $V[f^*_i,f_i]$ is parametrized by the many-body full vertex functions of 
the impurity problem defined by Eq.~\ref{eq:imp_action} (in practice, only the two-body vertex function 
is used)~\cite{Rubtsov08, hartmut_thesis}. In this way, the dual fermion lattice system is well-defined and thus provides sufficient 
input for a many-body diagrammatic calculation on the dual lattice.  After the dual lattice action of 
Eq.~\ref{eq:DFaction} is solved, the dual fermion Green function $G_d({\bf k},i\omega)$ is mapped back to the 
real fermion lattice via the relation of the form
\begin{equation}
G({\bf k},i\omega) = G_{imp}^{-2}(i\omega)(\Delta_w-\epsilon_{\bf k})^{-2}G_d({\bf k},i\omega)+(\Delta_w-\epsilon_{\bf k})^{-1}.
\label{Eq:realGF}
\end{equation}

This dual fermion formalism applies for both interacting and disordered~\cite{{h_terletska_13, df_afk, Rubtsov_disorder}} 
systems, provided that the dual potential is split into elastic and inelastic parts
and the closed fermion loops involving the elastic parts only are eliminated to prevent unphysical
renormalization of the interaction from scatterings from the disorder potential~\cite{h_terletska_13, df_afk}.

\subsection{Conventional Dual Fermion Algorithm}

\begin{figure}[tbh]
\centerline{ \includegraphics[clip,scale=0.6]{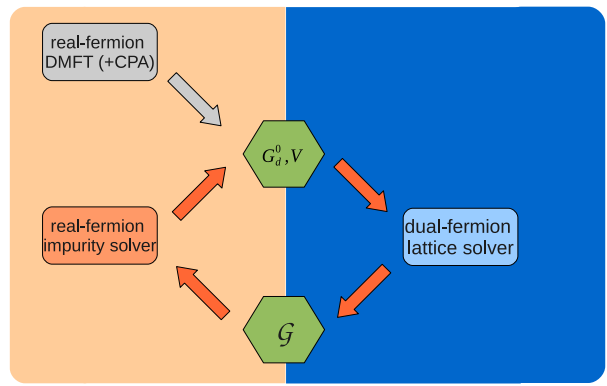}}

\caption{(Color online). Algorithm for the conventional dual fermion approach. The orange region (left half)
is for the real fermion impurity calculation, where the local on-site correlations are 
taken into account by quantum Monte Carlo (QMC), or other numerical methods. The blue region (right 
half) is for the dual fermion lattice calculation, where the nonlocal corrections ignored in 
the DMFT and/or CPA calculation are systematically restored.  The connection between these two regions 
is the dual fermion mapping.}

\label{fig:algorithm} %
\end{figure}

The conventional dual fermion algorithm is described in Fig.~\ref{fig:algorithm}. 
We start from the DMFT and/or CPA solution of the real fermion system, and then use the information
collected by solving the impurity problem (mainly the single-particle Green function $G_{imp}$,
self-energy $\Sigma_{imp}$, and two-particle
Green function $\chi_{imp}$) to parametrize the dual fermion system, i.e. to construct the bare 
dual fermion Green function $G^0_{d}$ and the dual potential $V[f^*,f ]$.
While the local correlations are described by the DMFT and/or CPA solution, the nonlocal corrections
are incorporated through the dual fermion part, which is calculated using standard perturbation
expansion in the $V$ term. 
After the dual fermion system is solved, we map it back to real fermion system with the nonlocal 
corrections included in the lattice self-energy $\Sigma({\bf k},i\omega)$ and  Green function $G({\bf k},i\omega)$.  We then 
solve the impurity problem again starting with an updated impurity-excluded Green function
$\mathcal{G}(i\omega)$. These steps are repeated until self-consistency is 
achieved with $ \sum_{\bf k} G_d({\bf k},i\omega) = 0$, i.e. with the local contribution to the dual fermion Green function
$G_d({\bf k},i\omega)$ being zero ~\cite{Rubtsov08}.

There are two predominantly time-consuming parts in the dual fermion calculation. One is the 
calculation of the two-particle Green function in the impurity or cluster solver, where the 
time needed is fixed for a given parameter set. The other is the solution of the dual fermion lattice
problem, where the time needed depends on the lattice size. Suppose the total system size is
$n_t = n_f \times L^D$ where $n_f$ is the number of frequencies used, $L$ is the linear lattice
size and $D$ is the dimension. The total number of sites in the lattice is $N_l=L^D$. 
Then the computational complexity of the dual fermion lattice 
calculation scales as
\begin{equation}
\mathcal{O}(n_t^2) = \mathcal{O}(n_f^2 \times L^{2D}) 
\end{equation}
for a second-order calculation,  
\begin{equation}
\mathcal{O}(n_t^3) = \mathcal{O}(n_f^3 \times L^{3D}) 
\end{equation}
for a fluctuation exchange (FLEX)~\cite{Bickers-flex} calculation, and
\begin{equation}
\mathcal{O}(n_t^4) = \mathcal{O}(n_f^4 \times L^{4D}) 
\end{equation}
for a two-particle self-consistent full parquet approach~\cite{c_dedominicis_64, yang_parquet}. To make sure that
the calculation is representative of the thermodynamic limit, the lattice linear size $L$ 
should be around 100 sites or larger. This imposes a severe constraint on the application of the 
dual fermion approach which so far has been applied only on one- and two-dimensional systems, and not yet 
on three-dimensional systems.  Even for one or two dimensions, the calculations are limited by the rapidly 
increasing computational complexity as the lattice size increases. 
Although the fast Fourier transform (FFT) might be used to reduce the computational complexity
to $\mathcal{O}(n_t \log_2(n_t))$ and $\mathcal{O}(n_t^2  \log_2(n_t))$ for the dual fermion second-order 
and FLEX calculations respectively, it is still very demanding when $L$ is large, 
and this reduction is not possible when using the parquet approach to solve the dual lattice problem.

Since the computational complexity depends on the linear size of the dual fermion lattice $L$, we would 
like to reduce that value as much as possible. In the conventional dual fermion approach, both the real 
fermion and dual fermion lattices share the same linear size $L$, so we would need to reduce the real fermion 
or the dual fermion system size.  Note that after solving the impurity problem, the dual fermion lattice system 
is well-defined via the bare dual Green function and bare dual potential. In this sense, there is no
difference as compared to the real fermion system. Thus, we can use any action-based approach available 
for the real fermion system to solve the dual fermion lattice problem.  Using a second-order perturbation 
theory or FLEX  for the conventional dual fermion approach can be interpreted as a finite-size calculation, 
and finite-size effects can be large.  If we want to eliminate or reduce these finite-size effects, we 
can embed our dual fermion calculation in an effective medium.  
In the following, we will propose two such embedding schemes.

\subsection{Real Fermion Embedding}

\begin{figure}[tbh]
\centerline{$\;\;\;\;\;\;$ 
            \includegraphics[clip,scale=0.5]{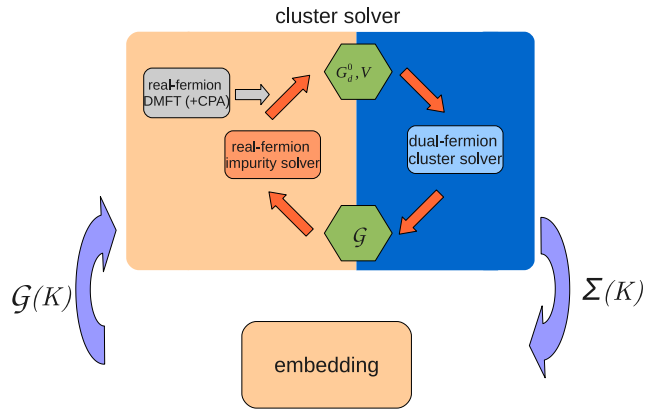}}

\caption{(Color online). Algorithm for the real fermion embedding scheme. It
is essentially the DCA algorithm with the dual fermion approach employed
as the cluster solver. The dual fermion mapping is implemented on the DCA 
cluster where the impurity is embedded. }

\label{fig:algorithm-rf} %
\end{figure}

In the first approach, which we refer to as real fermion embedding, we use the concepts of coarse graining introduced 
in the DCA~\cite{m_hettler_98a,m_hettler_00a} to map the real lattice to a cluster 
embedded in a self-consistently determined medium.  However, unlike in the conventional DCA, here the cluster problem 
is solved using the dual fermion method (see Fig.~\ref{fig:algorithm-rf}). Therefore, we employ the conventional dual fermion 
approach as the DCA cluster solver where the cluster size $L_c$ can be chosen to be small, of the order of 
several dozen sites, and the cluster is embedded in a self-consistently determined real fermion mean field.  
If any ${\bf {\bf k}}$ momentum on the lattice and the $N_c=L_c^D$ cluster momentum ${\bf K}$ are related as 
${\bf k}={\bf K}+\tilde{{\bf k}}$ with $\tilde{{\bf k}}$ labeling
the momentum within a coarse-graining cell surrounding ${\bf K}$, then the coarse graining sums over $\tilde{{\bf k}}$ 
are straightforward since the self-energy and irreducible vertices are assumed to be independent of $\tilde{{\bf k}}$.  
These sums may be completed in what is essentially the thermodynamic limit by a direct summation or, for single band
models, by defining a partial bare single particle density of states.  In either case the number of $\tilde{{\bf k}}$
points can be chosen to be sufficiently large so that the thermodynamic limit is guaranteed in this algorithm. Note that in this 
embedding scheme the mean-field lives on a real fermion lattice.  Therefore, after solving the cluster, any information 
collected from the dual fermion cluster should be mapped back to real fermion cluster. To be specific, the algorithm 
can be described as follows, where we suppress the explicit frequency dependence to simplify these expressions:
\begin{itemize}
\item Given the real fermion cluster self-energy $\Sigma_c({\bf K})$ 
which in the DCA scheme approximates the self-energy of the real lattice, we calculate the coarse-grained lattice
Green function through:
\begin{equation}
\bar{G}({\bf K})=\frac{N_{c}}{N_{l}}\sum_{\tilde{{\bf k}}}\frac{1}{i\omega + \mu - \epsilon_{{\bf K}+\tilde{{\bf k}}}
-\Sigma_c({\bf K})}.
\label{eq:RFE_Gbar}
\end{equation}
Then the cluster-excluded Green function is calculated by removing the cluster self-energy contribution
\begin{equation}
\mathcal{G}({\bf K})=[\bar{G}^{-1}({\bf K})+\Sigma_c({\bf K})]^{-1}.
\label{eq:RFE_Gscript}
\end{equation}
\end{itemize}

\begin{itemize}
\item With the calculated cluster-excluded Green function $\mathcal{G}({\bf K})$, the cluster problem is well-defined. 
The next step involves solving the cluster problem using a conventional dual fermion algorithm as the solver.
Since here the "lattice" for the conventional dual fermion approach is actually a cluster with linear size $L_c$,
which itself is embedded in a mean-field lattice, the original bare lattice Green function 
should be replaced accordingly by the cluster-excluded Green function: 
\begin{equation}
G^0(\bf k)=\frac{1}{i\omega + \mu - \epsilon_{\bf k}} \to \mathcal{G}({\bf K})
\end{equation}
in Eq.~\ref{eq:action}.
The parametrization of the dual fermion cluster problem is also affected with modified definition of the
bare dual fermion Green function of Eq.~$(\ref{Eq:dual-Green})$ as
\begin{equation}
G^0_{d}({\bf K})=\frac{1}{\mathcal{G}^{-1}({\bf K})-\Sigma_{imp}}-G_{imp}.
\label{eq:RFE_Gd0}
\end{equation}
Notice that here, as in the conventional dual fermion scheme, the input $G^0_{d}$ to the dual fermion 
loop is constructed from the solutions of the impurity problem with impurity Green function 
$G_{imp}$ and self-energy $\Sigma_{imp}$. 

After the cluster problem is solved, we obtain the cluster real fermion Green function $G({\bf K})$. The 
cluster self-energy then can be updated via the Dyson equation
\begin{equation}
\Sigma_c({\bf K}) = \mathcal{G}^{-1}({\bf K}) - G^{-1}({\bf K}). 
\end{equation}
\end{itemize}
We iterate these two steps until the difference between the self-energy from two consecutive iterations is below 
a given convergence criterion. Note that the real fermion cluster self-energy is used to approximate the 
lattice self-energy. For two-particle quantities, similarly, the real fermion irreducible vertex function 
is used to approximate the lattice irreducible vertex function and then the full vertex functions, 
two-particle Green functions and conductivity can be calculated accordingly~\cite{DCA_QMC}.

\subsection{Dual Fermion Embedding}

\begin{figure}[tbh]
\centerline{$\;\;\;\;\;\;\;\;\;\;\;$
            \includegraphics[clip,scale=0.6]{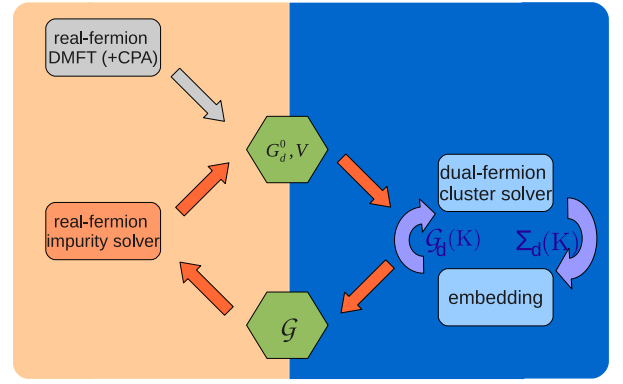}}

\caption{(Color online). Algorithm for the dual fermion embedding scheme. 
Similar to the conventional dual fermion approach, the dual fermion mapping is implemented at
the level of lattice. However, unlike in the conventional dual fermion scheme, the 
dual fermion lattice problem is solved using the DCA approach instead of a finite size calculation.}

\label{fig:algorithm-df} %
\end{figure}

As an alternative to reduce the computational complexity in the dual fermion lattice 
calculation, we employ the DCA-like scheme on the dual fermion lattice directly. We refer to
this approach as a dual-fermion embedding method, where the dual fermion lattice is replaced 
by a finite dual fermion cluster embedded in a self-consistently determined host.    The 
proposed dual fermion embedding algorithm is described in Fig.~\ref{fig:algorithm-df}.

The DCA algorithm for the dual fermion lattice is similar to the real fermion algorithm described 
above. Again taking the momentum ${\bf K}$ on a cluster of size $N_c$ and the ${\bf k}={\bf K}+\tilde{{\bf k}}$ 
on the lattice, we can write down the dual fermion embedding algorithm as follows: 
\begin{itemize}
\item Given the dual fermion cluster self-energy $\Sigma_{d}({\bf K})$ (either from an initial guess or from the previous 
iteration), we calculate the coarse-grained dual fermion lattice Green function $\bar{G}_{d}({\bf K})$ through 
\begin{equation}
\bar{G}_{d}({\bf K})=\frac{N_{c}}{N_{l}}\sum_{\tilde{{\bf k}}}\frac{1}{G_{d}^{0\,-1}
({\bf K}+\tilde{{\bf k}})-\Sigma_{d}({\bf K})},
\label{eq:DFE_Gbar}
\end{equation}
where the bare dual Green function is defined as
\end{itemize}

\begin{equation}
G^0_{d}({\bf K}+\tilde{{\bf k}})=\frac{1}{i\omega+\mu-\epsilon_{{\bf k}}-\Sigma_{imp}}-G_{imp}.
\end{equation}

\begin{itemize}
\item We then calculate the cluster-excluded dual fermion Green function $\mathcal{G}_{d}({\bf K})$ 
by removing the dual fermion 
cluster self-energy
\end{itemize}
\begin{equation}
\mathcal{G}_{d}({\bf K})=[\bar{G}^{-1}_{d}({\bf K})+\Sigma_{d}({\bf K})]^{-1}.
\label{eq:DFE_Gdscript}
\end{equation}

\begin{itemize}
\item This dual fermion cluster-excluded Green function $\mathcal{G}_{d}({\bf K})$ is the bare 
Green function on the dual fermion cluster, while the impurity full vertex is the bare dual interaction. 
Together, these two quantities define a perturbation theory that we may solve with various diagrammatic methods. 

As an example, 
if the self-consistent second-order theory is used, we will iterate the following two equations: 
\begin{equation}
G_{d}({\bf K})=[\mathcal{G}_{d}^{-1}({\bf K})-\Sigma_{d}({\bf K})]^{-1}
\end{equation}
and
\begin{eqnarray}
 &  & \Sigma_{d}(i\omega,{\bf K})\nonumber \\
 & = & -\frac{T^{2}}{N_{c}^{2}}\sum_{\omega^{\prime},\nu,{\bf K}^{\prime},{\bf Q}}
V_{i\omega,i\omega^{\prime},\nu}^{2}G_d(i\omega+i\nu,{\bf K}+{\bf Q})\nonumber \\
 &  & \;\;\;\;\;\;\;\;\;\;\;\;\times G_d(i\omega^{\prime}+i\nu,{\bf K}^{\prime}
+{\bf Q})G_d(i\omega^{\prime},{\bf K}^{\prime}),
\end{eqnarray}
until the self-consistency criterion for this inner loop is satisfied. 

We can also use a simplified FLEX algorithm in which the self-energy is calculated from ladder 
summations where all scattering channels are treated on a equal footing. We calculate the two-particle quantities 
after the self-energy has converged by rotating these ladder contributions into
the crossed channels using the parquet equations for the irreducible vertex functions. Details 
of the simplified FLEX method have been presented elsewhere~\cite{h_terletska_13, Janis_parquet}
and will not be discussed here.  

\end{itemize}

After the DCA loop
is converged and the dual lattice quantities are calculated, we continue as in the conventional 
dual fermion scheme, and  use the obtained dual fermion quantities  to parametrize their real 
lattice counterparts (e.g., Eq.~\ref{Eq:realGF}), and repeat the whole procedure until 
self-consistency is reached.

\section{Results for Anderson disorder model}
To qualify these new embedding schemes, we first apply them to the one-dimensional Anderson disorder model with the 
Hamiltonian 
\begin{equation} 
\mathcal{H}=-t\sum_{<ij>}c_{i}^{\dagger}c_{j}+\sum_{i}\epsilon_{i}n_{i},
\label{eq:AndersonH}
\end{equation}
where only the nearest neighbor hopping, $t$, is included, $4t=1$ sets the unit of
energy,  and the on-site disorder potential $\epsilon_i$ is distributed according to 
\begin{equation} 
\mathcal{P}(\epsilon_{i})=\Theta(V/2-|\epsilon_{i}|)/V
\end{equation}
where $\Theta(x)$ is the step function
\begin{equation} 
\Theta(x) = \left\{
\begin{array}{c l}
  1, & x \ge 0 \\
  0, & x < 0
\end{array}
\right.
\end{equation}
In the following, we will explore both single-particle and two-particle quantities using
the dual fermion embedding algorithms described in Figs.~\ref{fig:algorithm-rf} and ~\ref{fig:algorithm-df}.

\subsection{Comparison of the two embedding schemes}

\begin{table}[tph]
\centerline{ \begin{tabular}{|c|c|c|c|c|}
\hline 
T & V & RF embedding & DF embedding & conventional DF\tabularnewline
\hline
\hline 
0.05 & 1.0 & 4 & 2 & 2\tabularnewline
\hline 
0.05 & 2.0 & 4 & 2 & 2\tabularnewline
\hline 
0.01 & 1.0 & 6 & 2 & 2\tabularnewline
\hline 
0.01 & 2.0 & 5 & 2 & 2\tabularnewline
\hline 
0.005 & 1.0 & 9 & 2 & 2\tabularnewline
\hline 
0.005 & 2.0 & 7 & 3 & 3\tabularnewline
\hline
\end{tabular}}
\caption{Comparison of the number of times the impurity problem needs to be solved to converge the real-fermion (RF) 
embedding, dual-fermion (DF) embedding and conventional DF algorithms for different values of temperature $T$ 
and disorder strength $V$ of the Anderson disorder model (Eq.~\ref{eq:AndersonH}). Although both embedding schemes
produce the same result within convergent criterion, the DF embedding needs to solve the impurity problem a smaller 
number of times and thus serves as a better choice to implement the embedding. Note that in the conventional DF algorithm
the impurity problem is solved the same number of times as in the proposed DF embedding,
hence no additional computational cost is needed in such embedding scheme.
}
\label{table} %
\end{table}

Numerical tests show that, for most cluster sizes and within the convergence criterion, both the dual and real 
fermion embedding algorithms produce the same results for 
both single-particle and two-particle quantities.  This is because the two approaches share many
of the same features, including similar definitions of the impurity problem and the bare
dual fermion interaction extracted from it.  They differ mainly in the definition of the
bare dual fermion Green function $G^0_{d}({\bf K})$.  As can be seen from Eqs.~\ref{eq:RFE_Gbar},
\ref{eq:RFE_Gscript} and \ref{eq:RFE_Gd0} the bare dual Green function used in the 
real fermion embedding, $G^0_{d}({\bf K})$, is dressed by the real fermion cluster self-energy
 $\Sigma_c({\bf K})$, while from Eqs.~\ref{eq:DFE_Gbar} to \ref{eq:DFE_Gdscript} the 
bare dual Green function used in the dual fermion embedding algorithm is dressed by the dual 
fermion self-energy.  
Conceptually, these two self-energies differ in that the real 
fermion cluster self-energy includes both local and nonlocal single particle renormalization,
while the dual fermion self-energy includes only nonlocal single particle renormalization.
However, in both algorithms, the bare dual Green functions are formed from cluster-excluded
Green functions, Eqs.~\ref{eq:RFE_Gscript} and \ref{eq:DFE_Gdscript} to prevent overcounting 
of the cluster diagrams, so that these Green functions are bare on the local cluster.  So, 
at least conceptually, if not formally, the two bare Green functions contain the same 
information so that the two algorithms converge to nearly the same results.
 
However, the dual fermion embedding algorithm is a better choice. After the 
introduction of the embedding, the total time is generally dominated by the impurity 
solver, especially for the more realistic Hubbard-like model. 
The embedding in the real fermion scheme usually requires
additional iterations of the impurity solver to achieve convergence.  Table \ref{table} shows 
a comparison of the number of times the impurity problem needs to be solved to obtain convergence 
by the two embedding algorithms and the conventional dual fermion algorithm. 
Indeed, generally the real fermion embedding algorithm needs 2 to 4 more iterations 
of the impurity solver than the dual fermion one.  
We also want to emphasize that the dual fermion embedding algorithm does not incur in additional iterations 
for the outer loop as compared to the conventional DF approach and thus does not increase the number 
of times the impurity problem is solved. Therefore, in the following, 
we only show results calculated using the dual fermion embedding algorithm.

\subsection{System size dependence of the local Green function}

\begin{figure}[tbh]
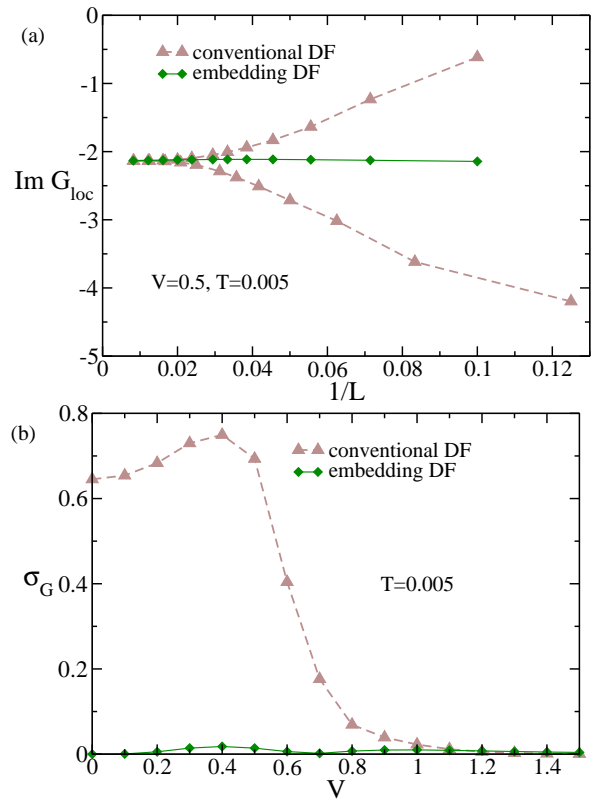

\centerline{ \includegraphics[clip,scale=0.3]{gl_V0.5_T0.005.eps}}

\centerline{\includegraphics[clip,scale=0.3]{gl_V_T0.005.eps}}

\caption{(Color online). 
Single-particle results for the one-dimensional Anderson disorder model at half-filling.
(a), the system size $L$ dependence of the imaginary part of the local Green function 
at the lowest Matsubara frequency $ImG_{loc}(i\pi T)$  for the conventional 
and the embedding dual fermion approximations for $V=0.5$ at temperature $T=0.005$ ($4t=1$). 
The conventional dual fermion calculation shows a large lattice 
size dependence, while the dual fermion embedding
calculation is almost flat as a function of the cluster size.
(b), the disorder strength $V$ dependence of the relative finite-size error $\sigma_G$ as defined in 
Eq.~\ref{eq:sigmaG}. This error is larger at small and intermediate disorder strengths where
the embedding DF helps most in reducing this finite-size effect. }

\label{fig:gl-Lx} %
\end{figure}

Since the dual fermion formalism is a Green function based approach, 
we can analyze finite-size effects by looking into the local Green function 
at the lowest Matsubara frequency point $i\omega_0=i\pi T$ ($N$ is the system size)
\begin{equation} 
G_{loc}(i\omega_0) = \frac{1}{N} \sum_{\bf{k}} G(i\omega_0, \bf{k}).
\label{local_GF}
\end{equation}
Fig.~\ref{fig:gl-Lx}(a) shows a comparison of results from both the conventional dual fermion and the dual 
fermion embedding algorithms at disorder strength $V=0.5$ and temperature $T=0.005$.
Results calculated from the conventional 
dual fermion approach oscillate and have a two-branch structure depending on whether $n$, where the linear system size
$L=2n$ ($N=L^{D}$ where $D$ is the dimension and here $D=1$), is an odd or even number. The linear system size $L$ 
has to be as large as 100
to achieve converged results. In contrast, the results from the embedding dual fermion algorithm
converge very quickly with increasing cluster size $L$ and form a nearly flat line for the values 
of $L$ plotted. In addition, the oscillation and two-branch structure are absent, perhaps due to the 
fast convergence.

Fig.~\ref{fig:gl-Lx}(b) shows the disorder strength $V$ dependence of the relative finite-size error
which can be described by the following quantity
\begin{equation} 
\sigma_G = \frac{Im G_{loc}(i\omega_0)|_{L=30} - Im G_{loc}(i\omega_0)|_{L=10}}{Im G_{loc}(i\omega_0)|_{L=30}} 
\label{eq:sigmaG}
\end{equation}
calculated for two linear cluster sizes $L=10$ and $L=30$. 
This error is maximum in the small and intermediate disorder region where
the DF embedding helps most in reducing this finite-size effect. For strong disorder ($V>1$), 
the finite-size effects are weak and thus there is no difference between the 
conventional DF and the embedding DF approaches.

\subsection{System size dependence of the conductivity}

\begin{figure}[tbh]
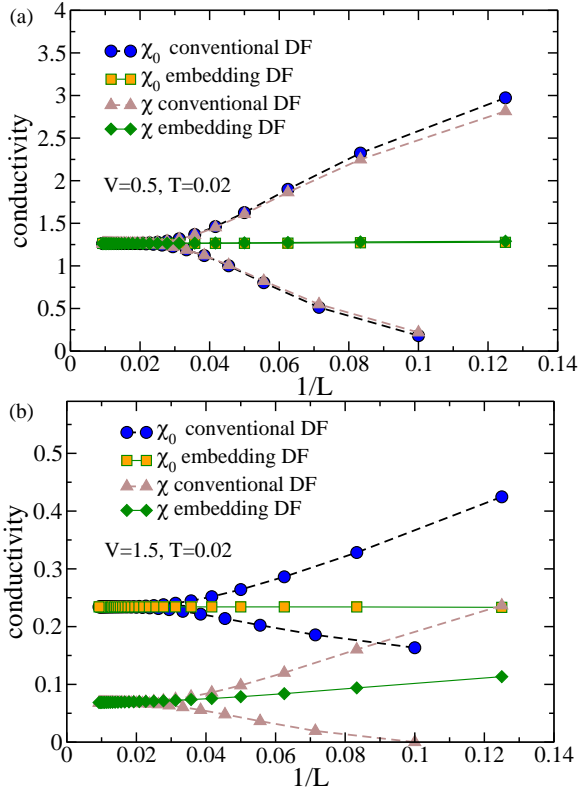

\centerline{ \includegraphics[clip,scale=0.3]{cond_V0.5_T0.02.eps}}

\centerline{\includegraphics[clip,scale=0.3]{cond_V1.5_T0.02.eps}}

\caption{(Color online). 
The system size dependence of the conductivity for the one-dimensional Anderson disorder model 
at half-filling from the conventional dual fermion
and the embedding dual fermion algorithms for $V=0.5$ (a) and $V=1.5$ (b) 
at temperature $T=0.02$. The conductivity has a larger size dependence 
as compared to the single-particle measurements. Nevertheless, the embedding scheme 
greatly reduces this size dependence.}

\label{fig:cond-Lx}%
\end{figure}

The second quantity we analyze is the dc conductivity $\sigma_{dc}$, which is
a two-particle quantity. At low temperatures, it can be approximated as \cite{{Scalettar},{Vollhardt-conductivity}} 
\begin{equation}
\sigma_{dc}=\frac{\beta^{2}}{\pi}\Lambda_{xx}{ \left( {\bf q } =0,\tau=\frac{\beta}{2}\right)},\label{dc_cond}
\end{equation}
where $\beta=1/k_BT$, and the current-current correlation function is $\Lambda_{xx}{\bf (q=0},\tau)=<j_x({\bf q},\tau)j_x(-{\bf q},0)>$. 
Such lattice correlation functions are obtained from the dual fermion two-particle 
Green function $\chi_d=-\chi^0_d-\chi^0_d F_d\chi^0_d$, with $\chi^0_d=G_d G_d$ ~\cite{Rubtsov08}. 
Here, the full dual fermion vertex  $F_d$ is obtained from the Bethe-Salpeter equation ~\cite{{Bickers}, {Janis_parquet}, {Hafermann}}
$F_d=\Gamma_d +\Gamma_d \chi^0_d F_d$. The conductivity hence can be decomposed into two parts, 
$\sigma=\sigma_0+\Delta \sigma$, where $\sigma_0$ is the mean-field Drude conductivity, coming from 
the bare bubble $\chi^0$, and the second part $\Delta \sigma$ incorporates the vertex corrections. 
 
Fig.~\ref{fig:cond-Lx} shows a comparison of the results. As compared to the single-particle 
quantities, the dependence of the conductivity on $L$ is much more severe. Nevertheless,
the embedding dual fermion method does a much better job on reducing this dependence. One interesting
observation is that the conductivity calculated with vertex corrections ($\chi$) has a larger 
dependence on $L$ than the one without vertex corrections ($\chi^0$), especially for large values of disorder.

\section{Results for Hubbard model}
To further exemplify the advantage of the new embedding technique, we apply it to 
the two-dimensional Hubbard model
\begin{equation} 
\mathcal{H}=-t\sum_{<ij>\sigma}c_{i\sigma}^{\dagger}c_{j\sigma}- \mu \sum_{i\sigma}n_{i\sigma} 
           + U \sum_{i}(n_{i\uparrow}-\frac{1}{2})(n_{i\downarrow}-\frac{1}{2}),
\label{eq:HubbardH}
\end{equation}
where only the nearest neighbor hopping $t$ is included ($4t=1$ sets the unit of
energy),  $\mu$ is the chemical potential, U is the on-site Coulomb interaction,
and $n_{i\uparrow}=c_{i\uparrow}^{\dagger}c_{i\uparrow}$.
In the following, we will explore the dual fermion cluster size dependence of the 
local Green function at both half-filling and off-half-filling.

\subsection{Half-filling}

\begin{figure}[tbh]
\centerline{ \includegraphics[clip,scale=0.32]{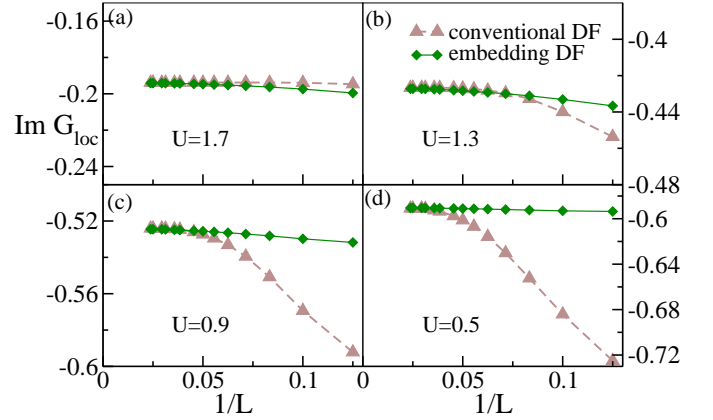}}
\caption{(Color online). 
The linear system size $L$ dependence of the imaginary part of the local Green function $ImG_{loc}$
for the conventional and the embedding dual fermion approaches for $T=0.025$ and 
different values of U's for the two-dimensional Hubbard model at half-filling. 
For the large U case, the finite-size effect is small, and both conventional and embedding dual fermion 
approximations converge quickly and produce similar results. With decreasing U, the finite-size effects
become more pronounced and the embedding dual fermion approach yields faster and more consistent
results.}
\label{fig:gl-U} %
\end{figure}

Fig.~\ref{fig:gl-U} shows the linear system size $L$ dependence of the imaginary part of local Green function 
(Eq.~\ref{local_GF}) for the conventional 
and the embedding dual fermion approximations for $T=0.025$ ($4t=1$) and different U's at half-filling. 
For large U, finite-size effects are small, and both conventional and embedding dual fermion 
approaches converge quickly and produce similar results. With decreasing U, finite-size effects
become more pronounced and embedding dual fermion approach yields faster and more consistent
results. This behavior is consistent with calculations on the real fermion lattice, 
where the convergence is enhanced when using embedding techniques, such as the DCA, when the system is in the metallic region.

\subsection{Off-half-filling}
\begin{figure}[tbh]
\centerline{ \includegraphics[clip,scale=0.32]{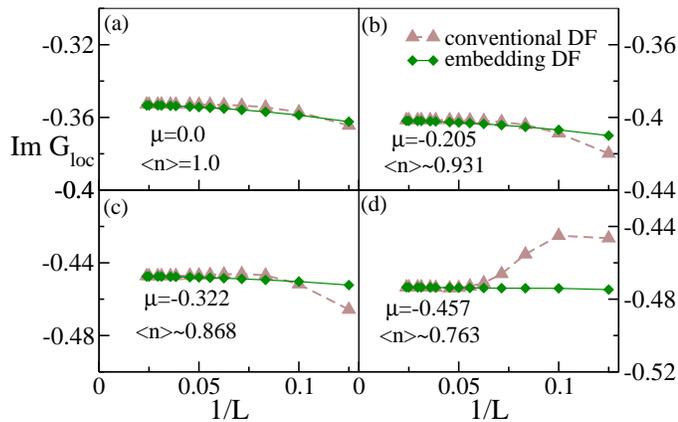}}
\caption{(Color online). 
The linear system size dependence of the imaginary part of the local Green function for the conventional 
and the embedding dual fermion calculations for $T=0.025$ ($4t=1$) and $U=1.5$ for the two-dimensional Hubbard model 
at different fillings $\langle n \rangle $.  
Similar to the situation of decreasing U at half-filling, doping the system away from half-filling tends to increase 
the finite-size effects. Embedding the dual fermion lattice helps considerably when finite-size effects are large, 
especially for the large doping case, $\langle n \rangle \sim 0.763$ of panel (d). }
\label{fig:gl-n} %
\end{figure}

Next we study the off-half-filling case. 
Fig.~\ref{fig:gl-n} shows the system size dependence of the imaginary part of the local Green function for the conventional 
and embedding dual fermion approaches for $T=0.025$ ($4t=1$) and $U=1.5$ at different 
chemical potentials. The converged fillings are also shown in each panel. 
Similarly to decreasing U at half-filling, doping the system away from half-filling tends to increase 
the finite-size effects. The embedding dual fermion approach helps considerably when finite-size effect are large,
especially for large doping case, say $\langle n \rangle \sim 0.763$ in panel (d)
where the system is in the metallic region. This behavior is consistent with the half-filling case.

\section{Discussion and Conclusions}

One significant drawback of the conventional dual fermion algorithm is the rapidly growing computational cost of the dual fermion lattice
calculation with increasing system size. This dependence is especially problematic if higher-order diagrammatic methods,
such as the FLEX or parquet approaches, are used to solve the dual fermion lattice problem.  The two embedding dual fermion schemes that we 
propose in this paper greatly reduce this computational cost. The first scheme, where the embedding is done on the real 
fermion lattice, is essentially the DCA method with the conventional dual fermion approach used as the cluster solver. As a general 
rule, any quantum method providing a good estimate of the single-particle Green function or self-energy can 
be employed in the DCA method as a cluster solver, and this embedding should help reduce the system size dependence of 
the solution.  In our second proposed embedding scheme, DCA coarse-graining method is applied directly to the dual fermion
lattice problem.  We find that this dual fermion embedding method provides much faster convergence with cluster size as 
compared to the convergence of the conventional dual fermion method with lattice size. This manipulation is possible because the 
dual fermion mapping defines an effective lattice system with a bare dual Green function and dual potential, and thus any 
action-based method useful for the real fermion system may also be employed in the dual fermion lattice calculation with only minor 
changes.

Our numerical tests show the real fermion and dual fermion embedding approaches converge to essentially the same result. 
However, the embedding in the dual fermion lattice turns out to be a much better choice since it requires a smaller number of 
iterations of the impurity solver.

The application of the embedding in the dual fermion lattice for the calculation of single-particle quantities for the Anderson disorder model
shows a faster convergence with system size as compared to the conventional dual fermion method, and the calculation of 
two-particle quantities also presents a large improvement of the convergence. 
And its application on the two-dimensional Hubbard model confirms the advantage of using the embedding technique in 
the dual fermion calculation for both half-filling and off-half-filing cases where finite-size effects are significant.

The proposed dual fermion embedding method should be even more advantageous in high-dimensional dual fermion calculations, especially in 
three dimensions. Only minimum changes are needed to introduce such a embedding in  current dual fermion codes.  By greatly reducing the
computational cost of the dual fermion diagrammatic calculations, these embedding schemes will also enable higher order 
approximations for the dual fermion diagrammatics, including potentially the full parquet approximation.

{\bf Acknowledgments.}
This work is supported by the DOE SciDAC grant DE-FC02-10ER25916 (SY and MJ) and BES CMCSN grant DE-AC02-98CH10886 (HT).  
Additional support was provided by NSF EPSCoR Cooperative Agreement No. EPS-1003897 (ZM and JM).

\appendix

\section{Dynamical Mean-Field Theory and Dynamical Cluster Approximation} 
For completeness, in this appendix we give a very brief introduction to the Dynamical Mean-Field Theory (DMFT)
and the Dynamical Cluster Approximation (DCA).
For a more detailed description, we refer interested readers to the vast literature available,
such as Refs.~\onlinecite{w_metzner_89a,e_mullerhartmann_89a,t_pruschke_95,Georges_DMFT} for the DMFT, 
and Refs.~\onlinecite{m_hettler_98a,m_hettler_00a,m_jarrell_01a,Maier05} for the DCA.

\subsection{Dynamical Mean-Field Theory}

\begin{figure}[tbh]
\centerline{ \includegraphics[clip,scale=0.5]{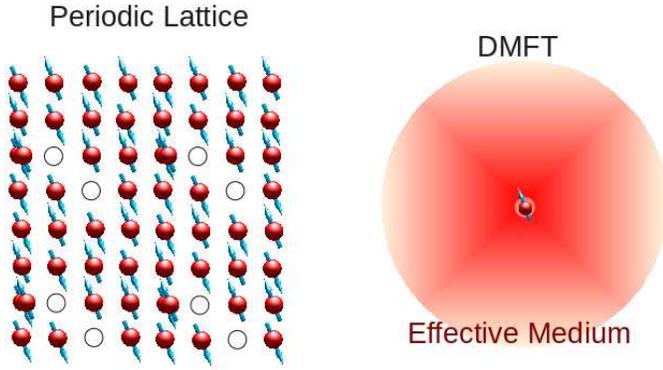}}
\caption{(Color online). 
Within DMFT, the original lattice model is mapped onto an impurity site embedded 
in a self-consistently determined effective mean-field medium.  }
\label{fig:dmfa} %
\end{figure}

It is usually very difficult to solve lattice models directly due to the exponential increase of the
computational costs with the system size because of the interdependent correlations at different length scales.
The philosophy behind the DMFT is to treat the local physics numerically exactly, while the non-local
fluctuations are treated at a mean-field level. In this way, as showed in Fig.~\ref{fig:dmfa},
the original lattice system is mapped onto an impurity site embedded in a self-consistently 
determined effective mean-field medium. This impurity system plus the mean field can be described 
by the Anderson impurity model, and many numerical methods are available to solve it. 
Since the mean field needs to be self-consistently determined, an iterative approach is best suited.
The algorithm is described in Fig.~\ref{fig:dmfa-algorithm}. Note that, as in the main text, 
we hide the explicit frequency dependence of each quantity to simplify the expressions in the 
following:
\begin{figure}[tbh]
\centerline{ \includegraphics[clip,scale=0.45]{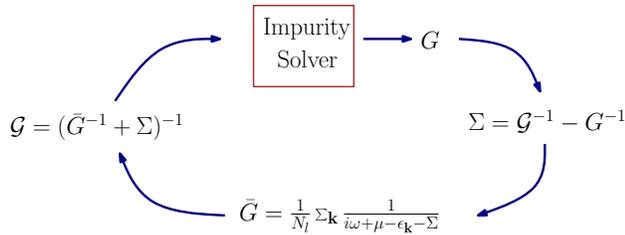}}
\caption{(Color online). 
DMFT algorithm. }
\label{fig:dmfa-algorithm} %
\end{figure}
\begin{itemize}
\item Given the initial impurity self-energy $\Sigma$ either from perturbation theory or
from a previous iteration, we calculate the coarse-grained lattice Green function through
\begin{equation}
\bar{G}=\frac{1}{N_{l}}\sum_{{\bf k}}\frac{1}{i\omega + \mu - \epsilon_{\bf k} - \Sigma}.
\end{equation}
Then the impurity-excluded Green function is calculated by removing the impurity self-energy 
contribution
\begin{equation}
\mathcal{G} = [\bar{G}^{-1}+\Sigma]^{-1}.
\end{equation}
\end{itemize}
\begin{itemize}
\item With the calculated impurity-excluded Green function $\mathcal{G}$, the impurity problem 
is well-defined.   After the impurity problem is solved, the obtained impurity Green function 
$G$ is used to update the impurity self-energy via the Dyson equation
\begin{equation}
\Sigma_c = \mathcal{G}^{-1} - G^{-1}. 
\end{equation}
\end{itemize}
These two steps are iterated until the convergence criterion is satisfied.

\subsection{Dynamical Cluster Approximation}

The DMFT is best suited for studying the local physics, e.g., Mott physics. 
However, as a single-site approximation it neglects non-local correlations, and hence can not
capture the non-local physics, e.g., d-wave superconductivity. 
To deal with this deficiency of the DMFT, cluster extensions, such as the DCA, have been proposed. 
\begin{figure}[tbh]
\centerline{\includegraphics[clip,scale=0.43]{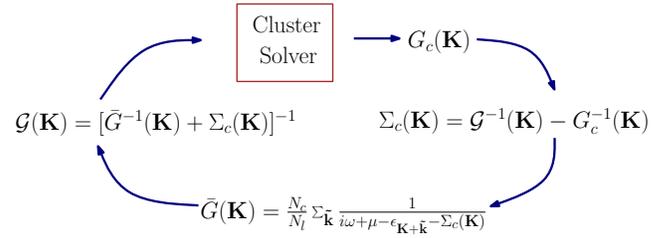}}
\caption{(Color online). DCA algorithm.  }
\label{fig:dca-algorithm} %
\end{figure}

Within the DCA, the original lattice system is mapped onto a periodic cluster (containing multiple 
sites) instead of an impurity site, embedded in a self-consistently deterrmined mean field. Now, the calculated quantities 
acquire cluster momentum $\bf K$ dependence. As depicted in Fig.~\ref{fig:dca-algorithm}, the algorithm 
can be described as:
\begin{itemize}
\item Given the initial cluster self-energy $\Sigma_c({\bf K})$ either from perturbation theory or
from a previous iteration, we calculate the coarse-grained lattice Green function through
\begin{equation}
\bar{G}({\bf K})=\frac{N_{c}}{N_{l}}\sum_{\tilde{{\bf k}}}\frac{1}{i\omega + \mu - \epsilon_{{\bf K}+\tilde{{\bf k}}}
-\Sigma_c({\bf K})}.
\end{equation}
Then the cluster-excluded Green function $\mathcal{G}({\bf K})$ is calculated by removing the 
cluster self-energy contribution
\begin{equation}
\mathcal{G}({\bf K})=[\bar{G}^{-1}({\bf K})+\Sigma_c({\bf K})]^{-1}.
\end{equation}
\end{itemize}
\begin{itemize}
\item With the calculated cluster-excluded Green function $\mathcal{G}({\bf K})$, the cluster 
problem is well-defined.  It can be solved by different numerical cluster solvers yielding the 
cluster Green function $G_c({\bf K})$.  The cluster self-energy then can be updated via the 
Dyson equation
\begin{equation}
\Sigma_c({\bf K}) = \mathcal{G}^{-1}({\bf K}) - G^{-1}_c({\bf K}). 
\end{equation}
\end{itemize}
These two steps are iterated until the convergence criterion is satisfied.

\bibliography{df_embedding}

\end{document}